\documentclass[preprint,amsmath,amssymb,aps]{revtex4}
\usepackage{graphicx}
\usepackage{bm}
\usepackage{subfigure}
\usepackage{amsmath}
\usepackage[utf8]{inputenc}

\newcommand{\pr}{Phys. Rev.\ }

\newcommand{\jpa}{J. Phys. A\ }

\newcommand{\etal}{{\em et al. }}
\newcommand{\etals}{{\em et al.}}  

\begin{document}

\title{Controlled asymmetry of EPR steering with an injected non-degenerate optical parametric oscillator}

\author{M.~K. Olsen}
\affiliation{School of Mathematics and Physics, University of Queensland, Brisbane, 
Queensland 4072, Australia.\\
Quantum Science Otago and Dodd-Walls Centre for Photonic and Quantum Technologies, Department of Physics, University of Otago, Dunedin, New Zealand.}

\date{\today}

\begin{abstract}

We propose and analyse a nonlinear optical apparatus in which the direction of asymmetric steering is controllable within the apparatus, rather than by adding noise to measurements. Using a nondegenerate parametric oscillator with an injected signal field, we show how the directionality and extent of the steering can be readily controlled for output modes which can be up to one octave apart. The two downconverted modes, which exhibit the greater violations of the steering inequalities, can also be controlled to exhibit asymmetric steering in some regimes.  

\end{abstract}

\maketitle


The existence of the phenomenon of steering was recognised by Schrödinger in 1935~\cite{Erwin1,Erwin2} as an extension of the Einstein Podolsky Rosen
paradox (EPR)~\cite{EPR}, and put on a firm mathematical footing by Wiseman \etal in 2007~\cite{Wisesteer}. The importance of EPR steering was reflected in a special journal issue on the topic~\cite{josabspecial}, and it has been shown to be necessary for secure continuous-variable teleportation~\cite{HeSteer}, with control being possible by feedback~\cite{Ajsad}.
Wiseman \etal also raised the question of whether asymmetric steering were possible; i.e. whether bipartite states shared between Alice and Bob existed where Alice could steer Bob, but not vice versa. For the case of Gaussian measurements, this was soon answered, both theoretically~\cite{SFG,sapatona} and experimentally~\cite{Natexp}, using the Reid EPR criteria for the products of inferred variances~\cite{EPRMDR}. It has since been established that asymmetric steering is a general property, not being dependent on Gaussian measurements~\cite{oneway}. Continuous variable asymmetric steering has been predicted in intracavity second harmonic generation~\cite{meu}, atomic Bose-Hubbard chains~\cite{CTAPJOSAB,BHchain}, and measured experimentally in a four-mode cluster state~\cite{Deng}. 

The optical parametric oscillator (OPO), along with homodyne measurements of phase-sensitive correlations, are mature technologies, found in many quantum optics laboratories~\cite{Hans}. The first experimental demonstration of EPR steering was by Ou \etals~\cite{Ou}, using the two downconverted fields of a non-degenerate OPO. An injected signal at one of the low frequency modes can be used to increase conversion efficiency as well as create a coherent component of the modes, in both the degenerate~\cite{degenOPO} and non-degenerate cases~\cite{ndegenOPO}. Yu \etal have shown how the non-degenerate OPO (NDOPO) can be used to produce three-colour entanglement~\cite{Yu}, with two of the same authors analysing the bichromatic entanglement properties with an injected signal~\cite{YuWang}. These bichromatic entanglement properties were experimentally investigated by Gu \etals~\cite{Guo}. Bichromatic entanglement was also analysed theoretically and experimentally with injected fields at both signal and idler by Wang and Li~\cite{WangLi}. In this work we show how controlling the amplitude of an injected signal can also control the asymmetry of EPR steering in the system. We examine EPR steering in all three possible output bipartitions and show that the effects are intrinsic to the scheme, not requiring added noise to achieve control of the quantum correlations as in previous work~\cite{Natexp,Qin}.  


The NDOPO consists of a nonlinear $\chi^{(2)}$ material inside a pumped Fabry-Perot cavity. Three optical fields interact inside the material: an externally pumped mode at frequency $\omega_{0}$, and two downconverted modes at $\omega_{1}$ and $\omega_{2}$, where $\omega_{0}=\omega_{1}+\omega_{2}$. The important aspect of the non-degeneracy is that the two down converted modes be distinguishable, so that they need not have different frequencies if they can be separated due to different polarisations, for example. In the system we examine here, we will consider the effects of an injected coherent signal at frequency $\omega_{1}$. Since the pump laser is often a high frequency mode from an upconversion process, a field at one of the lower frequencies should be readily available. 

The rotating wave interaction Hamiltonian for the system is
\begin{equation}
{\cal H}_{int} = i\hbar\kappa\left( \hat{a}_{0}\hat{a}_{1}^{\dag}\hat{a}_{2}^{\dag}-\hat{a}_{0}^{\dag}\hat{a}_{1}\hat{a}_{2}\right),
\label{eq:intham}
\end{equation}
where $\hat{a}_{j}$ is the bosonic annihilation operator for the mode at $\omega_{j}$ and $\kappa$ represents the effective $\chi^{(2)}$ nonlinearity.
The cavity pumping Hamiltonian is
\begin{equation}
{\cal H}_{pump} = i\hbar\left(\epsilon_{0}\hat{a}_{0}^{\dag}+\epsilon_{1}\hat{a}_{1}^{\dag} \right) + h.c,
\label{eq:Hpump}
\end{equation}
where the $\epsilon_{j}$ represent coherent input fields at frequency $\omega_{j}$. Note that we are considering that all fields are resonant with the cavity.
The damping of the cavity fields into a zero temperature Markovian reservoir is described by the Lindblad superoperator 
\begin{equation}
{\cal L}\rho = \sum_{i=0}^{2}\gamma_{i}\left(2\hat{a}_{i}\rho\hat{a}_{i}^{\dag}-\hat{a}_{i}^{\dag}\hat{a}_{i}\rho-\rho\hat{a}_{i}^{\dag}\hat{a}_{i} \right),
\label{eq:Lindblad}
\end{equation}
where $\rho$ is the system density matrix and $\gamma_{i}$ is the cavity loss rate at $\omega_{i}$.

Starting with the Hamiltonian, we proceed via the von Neumann equation for the density matrix, mapping this onto a Fokker-Planck equation (FPE) for the chosen pseudoprobability distribution, and then onto stochastic differential equations~\cite{QNoise}.
Since it is well known that the FPE for the Glauber-Sudarshan P-function~\cite{Roy,Sud} has a negative diffusion matrix and therefore cannot be mapped onto stochastic differential equations, we decide to use the positive-P distribution~\cite{P+}, which is exact for this system. This distribution requires a doubled phase space and the FPE can be simply found from the equation for the Glauber-Sudarshan P-distribution by setting variables and their complex conjugates as independent~\cite{Danbook}. This entails changing $\alpha_{j}^{\ast}$ to $\alpha_{j}^{+}$, so that $\alpha_{j}$ and $\alpha_{j}^{+}$ are now independent variables and allows for a positive-definite diffusion matrix in the resulting FPE. 

The resulting FPE is found as
\begin{eqnarray}
\frac{dP}{dt} &=& \left\{-\left[ \frac{\partial}{\partial\alpha_{0}}\left(\epsilon_{0}-\gamma_{0}\alpha_{0}-\kappa\alpha_{1}\alpha_{2} \right)+
\frac{\partial}{\partial\alpha_{0}^{+}}\left(\epsilon_{0}^{\ast}-\gamma_{0}\alpha_{0}^{+}-\kappa\alpha_{1}^{+}\alpha_{2}^{+} \right)\right.\right. \nonumber \\
& & \left.\left. +\frac{\partial}{\partial\alpha_{1}}\left(\epsilon_{1}-\gamma_{1}\alpha_{1}+\kappa\alpha_{0}\alpha_{2}^{+}\right)
+ \frac{\partial}{\partial\alpha_{1}^{+}}\left(\epsilon_{1}^{\ast}-\gamma_{1}\alpha_{1}^{+}+\kappa\alpha_{0}^{+}\alpha_{2}\right)
\right.\right. \nonumber \\
& & \left. \left. +\frac{\partial}{\partial\alpha_{2}}\left(-\gamma_{2}\alpha_{2}+\kappa\alpha_{0}\alpha_{1}^{+} \right)
+\frac{\partial}{\partial\alpha_{2}^{+}}\left(-\gamma_{2}\alpha_{2}^{+}+\kappa\alpha_{0}^{+}\alpha_{1}\right)\right]
\right. \nonumber \\
& & \left.+\frac{1}{2}\left( 2\frac{\partial^{2}}{\partial\alpha_{1}\partial\alpha_{2}}\kappa\alpha_{0}+ 2\frac{\partial^{2}}{\partial\alpha_{1}^{+}\partial\alpha_{2}^{+}}\kappa\alpha_{0}^{+}\right)\right\}P(\tilde{\alpha},t),
\label{eq:FPE}
\end{eqnarray}
where $\tilde{\alpha}$ is the vector of amplitude variables. This FPE maps onto six coupled stochastic differential equations,
\begin{eqnarray}
\frac{d\alpha_{0}}{dt} &=& \epsilon_{0}-\gamma_{0}\alpha_{0}-\kappa\alpha_{1}\alpha_{2}, \nonumber \\
\frac{d\alpha_{0}^{+}}{dt} &=& \epsilon_{0}^{\ast}-\gamma_{0}\alpha_{0}^{+}-\kappa\alpha_{1}^{+}\alpha_{2}^{+}, \nonumber \\
\frac{d\alpha_{1}}{dt} &=& \epsilon_{1}-\gamma_{1}\alpha_{1}+\kappa\alpha_{0}\alpha_{2}^{+}+\sqrt{\frac{\kappa\alpha_{0}}{2}}(\eta_{1}+i\eta_{2}), \nonumber \\
\frac{d\alpha_{1}^{+}}{dt} &=& \epsilon_{1}^{\ast}-\gamma_{1}\alpha_{1}^{+}+\kappa\alpha_{0}^{+}\alpha_{2}+\sqrt{\frac{\kappa\alpha_{0}^{+}}{2}}(\eta_{3}+i\eta_{4}), \nonumber \\
\frac{d\alpha_{2}}{dt} &=& -\gamma_{2}\alpha_{2}+\kappa\alpha_{0}\alpha_{1}^{+}+\sqrt{\frac{\kappa\alpha_{0}}{2}}(\eta_{1}-i\eta_{2}), \nonumber \\
\frac{d\alpha_{2}^{+}}{dt} &=& -\gamma_{2}\alpha_{2}^{+}+\kappa\alpha_{0}^{+}\alpha_{1}+\sqrt{\frac{\kappa\alpha_{0}^{+}}{2}}(\eta_{3}-i\eta_{4}),
\label{eq:PPNDOPO}
\end{eqnarray}
where the complex variable pairs $(\alpha_{i},\alpha_{j}^{+})$ correspond to the operator pairs $(\hat{a}_{i},\hat{a}_{j}^{\dag})$ in the sense that stochastic averages of products converge to normally-ordered operator expectation values, e.g. $\overline{\alpha_{i}^{+\,m}\alpha_{j}^{n}} \rightarrow \langle \hat{a}_{i}^{\dag\,m}\hat{a}_{j}^{n} \rangle$. The $\eta_{j}$ are Gaussian noise terms with the properties $\overline{\eta_{i}}=0$ and $\overline{\eta_{j}(t)\eta_{k}(t')}=\delta_{jk}\delta(t-t')$. We note that these equations have the same form in either It\^o or Stratonovich calculus~\cite{SMCrispin} and that they describe the process inside the optical cavity.


When nonlinear optical media are held inside a pumped optical cavity, the measured observables are usually the output spectral correlations, which are accessible using homodyne measurement techniques~\cite{mjc}. These are readily calculated in the steady-state by treating the system as an Ornstein-Uhlenbeck process~\cite{SMCrispin}. In order to do this, we begin by expanding the positive-P variables into their steady-state expectation values plus delta-correlated Gaussian fluctuation terms, e.g.
\begin{equation}
\alpha_{ss} \rightarrow \langle\hat{a}\rangle_{ss}+\delta\alpha.
\label{eq:fluctuate}
\end{equation}
Given that we can calculate the $\langle\hat{a}\rangle_{ss}$, we may then write the equations of motion for the fluctuation terms. The resulting equations are written for the vector of fluctuation terms as
\begin{equation}
d\delta\vec{\alpha} = -A\delta\vec{\alpha}dt+Bd\vec{W},
\label{eq:OEeqn}
\end{equation}
where $A$ is the steady-state drift matrix, $B$ is found from the factorisation of the diffusion matrix of the original Fokker-Planck equation, $D=BB^{T}$, with the steady-state values substituted in, and $d\vec{W}$ is a vector of Wiener increments. As long as the matrix $A$ has no eigenvalues with negative real parts, this method may be used to calculate the intracavity spectra via
\begin{equation}
S(\omega) = (A+i\omega\openone)^{-1}D(A^{\mbox{\small{T}}}-i\omega\openone)^{-1},
\label{eq:Sin}
\end{equation}
from which the output spectra are calculated using the standard input-output relations~\cite{mjc} and $\openone$ is the $6\times 6$ identity matrix. Note that the procedure for obtaining the matrix $S(\omega)$ by Fourier transform of the two-time covariance matrix is fully covered in Ref.~\cite{SMCrispin}, having been originally developed for stochastic analysis of chemical reactions by Chaturvedi \etals~\cite{OSChat}. 

In this case the semi-classical equations found by removing the noise terms from Eq.~\ref{eq:PPNDOPO} are difficult to solve analytically, requiring the solution of a fifth-order polynomial. For this reason, we will proceed numerically in what follows. $A$ is found as 
\begin{equation}
A =
\begin{bmatrix}
\gamma_{0} & 0 & \kappa\alpha_{2} & 0 & \kappa\alpha_{1} & 0 \\
0 & \gamma_{0} & 0 & \kappa\alpha_{2}^{\ast} & 0 & \kappa\alpha_{1}^{\ast} \\
-\kappa\alpha_{2}^{\ast} & 0 & \gamma_{1} & 0 & 0 & -\kappa\alpha_{0} \\
0 & -\kappa\alpha_{2} & 0 & \gamma_{1} & -\kappa\alpha_{0}^{\ast} & 0 \\
-\kappa\alpha_{1}^{\ast} & 0 & 0 & -\kappa\alpha_{0} & \gamma_{2} & 0 \\
0 & -\kappa\alpha_{1} & -\kappa\alpha_{0}^{\ast} & 0 & 0 & \gamma_{2} 
\end{bmatrix},
\label{eq:Amat}
\end{equation}
and $D$ is a $6\times 6 $ matrix with
\begin{eqnarray}
D(3,5) &=& D(5,3) = \kappa\alpha_{0}, \nonumber \\
D(4,6) &=& D(6,4) = \kappa\alpha_{0}^{\ast},
\label{eq:Dmat}
\end{eqnarray}
and all other elements being zero.
In the above two equations, the $\alpha_{j}$ should be read as the steady-state mean values, so that $\alpha_{j}^{\ast}=\overline{\alpha_{j}^{+}}$. These are now complex numbers which are the averages of the positive-P stochastic variables.
Because we have parametrised our system using $\gamma_{1}=1$, the frequency $\omega$ is in units of $\gamma_{1}$.
$S(\omega)$ is now in terms of quadratic products of the fluctuation operators. To express it in terms of the canonical quadratures, we calculate
\begin{equation}
S^{q}(\omega) = QSQ^{T},
\label{eq:quadratures}
\end{equation} 
where $Q$ is the block diagonal $6\times 6$ matrix constructed from
\begin{equation}
q =
\begin{bmatrix}
1 & 1 \\
-i & i
\end{bmatrix}.
\label{eq:qmat}
\end{equation}
$S^{q}(\omega)$ then gives us the products we require to construct the output spectral variances and covariances for modes $i$ and $j$ as, for example,
\begin{equation}
V(X_{i},X_{j}) = \delta_{ij}+\sqrt{\gamma_{i}\gamma_{j}} \left(S_{2i-1,2j-1}^{q}+S_{2j-1,2i-1}^{q}\right).
\label{eq:Sout}
\end{equation}
It is important to note here that this process is not valid if the eigenvalues of $A$ have any negative real parts, which is not the case for any of the results presented.


In order to show EPR steering, we use the Reid criterion~\cite{EPRMDR}, for which the product of two inferred quadrature variances being less than one proves the existence of the EPR paradox for that particular bipartition. The inferred variances are found as
\begin{eqnarray}
V_{inf}(\hat{X}_{ij}) &=& V(\hat{X}_{i})-\frac{\left[V(\hat{X}_{i},\hat{X}_{j})\right]^{2}}{V(\hat{X}_{j})}, \nonumber \\
V_{inf}(\hat{Y}_{ij}) &=& V(\hat{Y}_{i})-\frac{\left[V(\hat{Y}_{i},\hat{Y}_{j})\right]^{2}}{V(\hat{Y}_{j})},
\label{eq:VXYinf}
\end{eqnarray}
where $V(AB) = \langle AB\rangle-\langle A\rangle\langle B\rangle$ and $V_{inf}(A_{ij})$ denotes the variance of $A_{i}$ as inferred by measurements made of $A_{j}$. 
When the product of these two inferred variances is less than one, mode $i$ can be steered by measurements made at mode $j$, and the EPR paradox is demonstrated for these two modes.
We will use EPR$_{jk}$ as the product of the $\hat{X}_{jk}$ and $\hat{Y}_{jk}$ inferred variances. The directionality of the paradox is recognised in the fact that EPR$_{jk}$, where mode $j$ is steered by measurements of mode $k$, is not always equal to EPR$_{kj}$. The situation where one of these is less than one while the other is more than one is known as Gaussian asymmetric steering. We note that our quadrature definitions are $\hat{X}_{j}=\hat{a}_{j}+\hat{a}_{j}^{\dag}$ and $\hat{Y}_{j}=-i(\hat{a}_{j}-\hat{a}_{j}^{\dag})$. Because the EPR steerable states are a strict subset of the entangled states, both symmetric and asymmetric steering demonstrate that the two modes concerned are fully bipartite entangled.

\begin{figure}[tbhp]
\includegraphics[width=0.75\columnwidth]{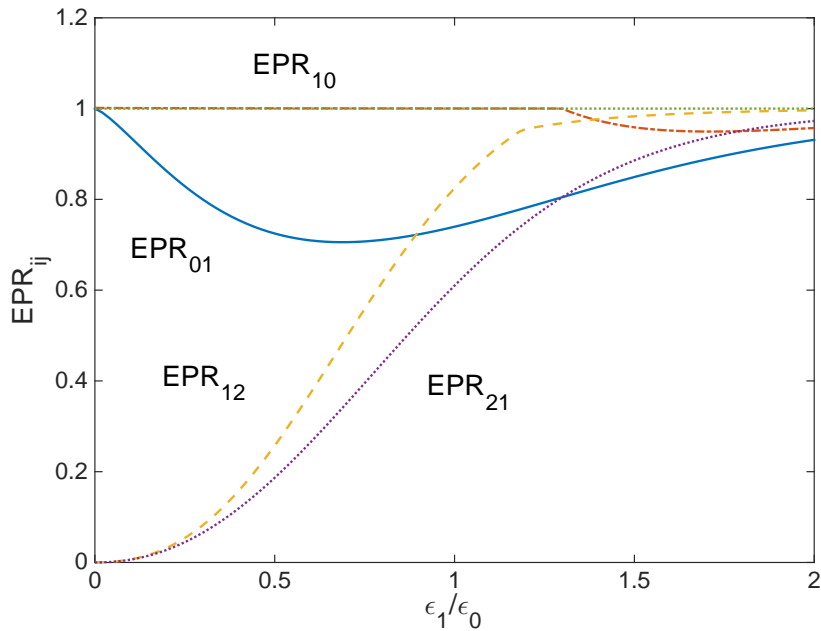}
\caption{(colour online) The minima of the spectral EPR$_{ij}$  output correlations between the modes $0$ and $1$, and $1$ and $2$, as a function of the ratio of the injected signal to the pump amplitude at $\omega_{0}$. $\gamma_{0}=\gamma_{1}=\gamma_{2}=1$ for this result, $\kappa=10^{-2}$, and $\epsilon_{0}=100$. All quantities plotted in this work are dimensionless.}
\label{fig:EPRgequal}
\end{figure}

We find that the presence or otherwise of asymmetric steering between the three output modes can be controlled by the simple mechanism of altering the amplitude of $\epsilon_{1}$, the injected signal. This can be seen in Fig.~\ref{fig:EPRgequal}, where EPR$_{01}$ is less than one across the whole range shown, while EPR$_{10}$ only drops below one for $\epsilon_{1}\gtrsim 1.28\epsilon_{0}$. For this result the mirror loss rates at all frequencies are equal. Controlling the signal amplitude is perhaps the simplest change that can be made to a non-degenerate parametric oscillator, and should be easier than dynamically changing mirror reflectivities or detunings. A large degree of symmetric violation of the Reid inequalities for the two down converted modes is available across much of the range for these parameters, with asymmetric steering only appearing when the actual steering is negligible. We did not find any steering involving the pair of fields at $\omega_{0}$ and $\omega_{2}$, for the whole parameter range investigated. 

\begin{figure}[tbhp]
\includegraphics[width=0.75\columnwidth]{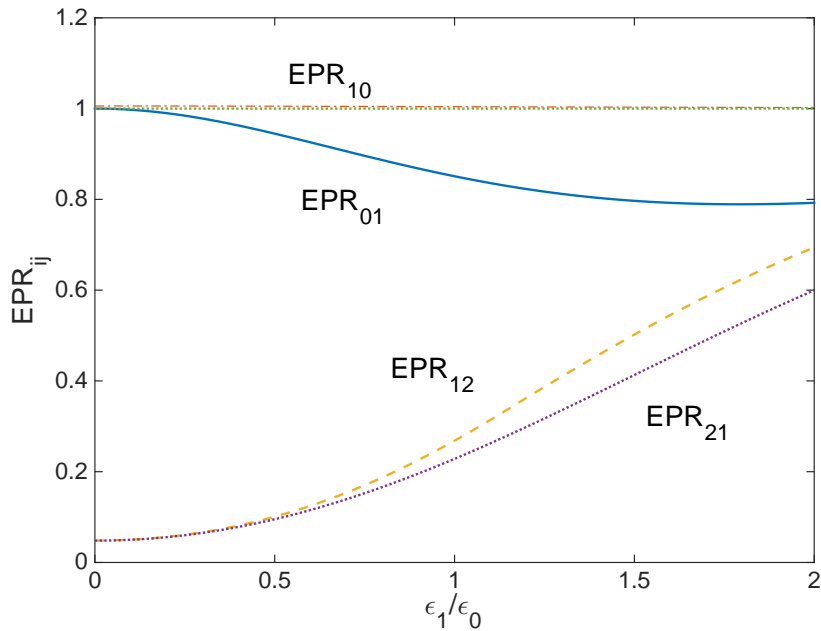}
\caption{(colour online) The minima of the spectral EPR$_{ij}$  output correlations between the modes $0$ and $1$, and $1$ and $2$, as a function of the ratio of the injected signal to the pump amplitude at $\omega_{0}$. $\gamma_{1}=\gamma_{2}=2=2\gamma_{0}$ for this result, $\kappa=10^{-2}$, and $\epsilon_{0}=100$.}
\label{fig:EPRg1g2two}
\end{figure}

It is also worthwhile to investigate the effects of different cavity loss rates on these phenomena. In practice, mirror losses can be either frequency dependent or polarisation dependent. When we double the loss rates for the downconverted modes, while leaving that at $\omega_{0}$ unchanged, we see no change from symmetric to asymmetric over the range of signal investigated. As shown in Fig.~\ref{fig:EPRg1g2two}, the pair $(1,2)$ exhibits symmetric steering across the whole range, while $(0,1)$ exhibits asymmetric steering.

 \begin{figure}[tbhp]
\includegraphics[width=0.75\columnwidth]{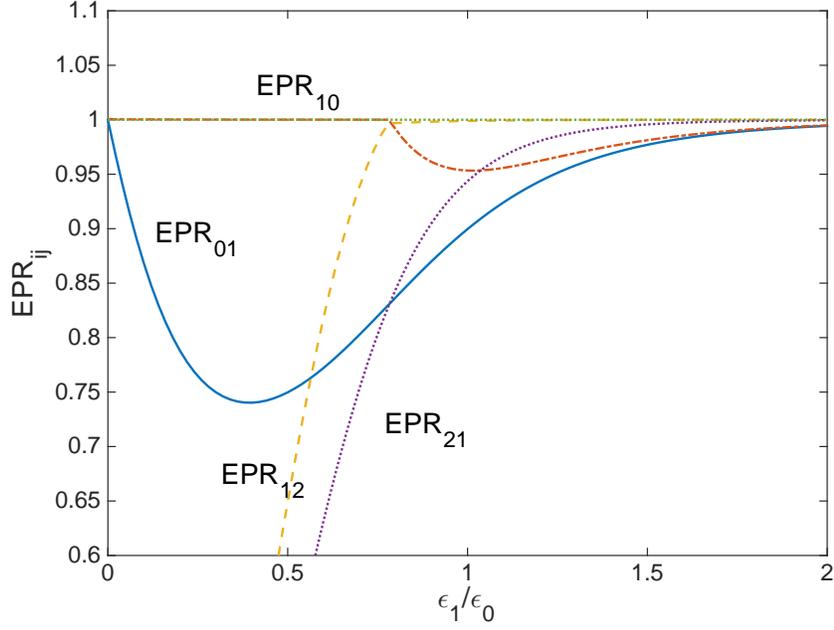}
\caption{(colour online) The minima of the spectral EPR$_{ij}$  output correlations between the modes $0$ and $1$, and $1$ and $2$, as a function of the ratio of the injected signal to the pump amplitude at $\omega_{0}$. $\gamma_{2}=\gamma_{0}=1=2\gamma_{1}$ for this result, $\kappa=10^{-2}$, and $\epsilon_{0}=100$.}
\label{fig:EPRg1half}
\end{figure} 

\begin{figure}[tbhp]
\includegraphics[width=0.75\columnwidth]{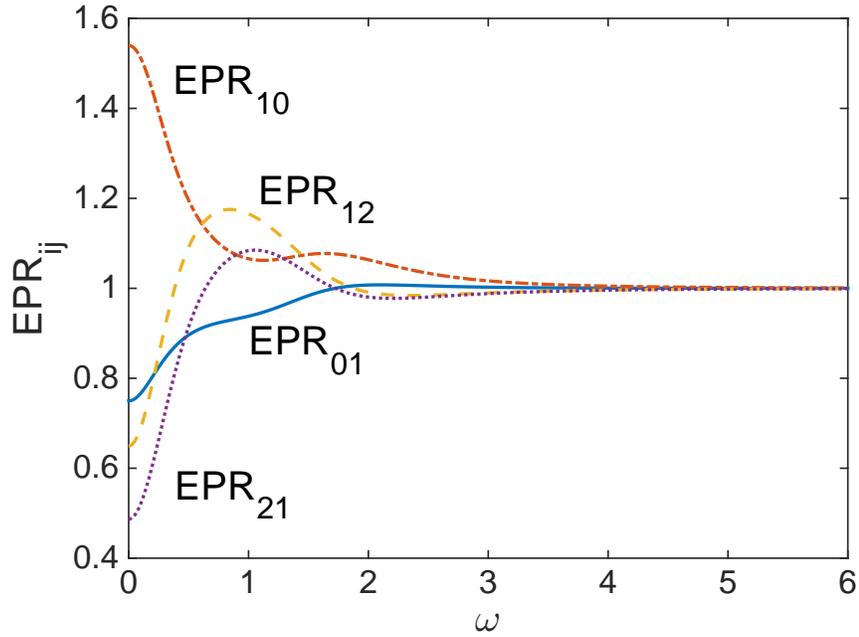}
\caption{(colour online) The positive frequency spectral EPR$_{ij}$  output correlations between the modes $0$ and $1$, and $1$ and $2$, for $\epsilon_{1}=50$. $\gamma_{2}=\gamma_{0}=1=2\gamma_{1}$ for this result, $\kappa=10^{-2}$, and $\epsilon_{0}=100$.}
\label{fig:spectra}
\end{figure}

A different example is the case where the injected field experiences a lower damping rate, as shown in Fig.~\ref{fig:EPRg1half}, where $\gamma_{0}=\gamma_{2}=1=2\gamma_{1}$. In this case we see a clear crossover for both bipartitions, at $\epsilon_{1}\approx 0.78\epsilon_{0}$, with $(0,1)$ being asymmetric below this, and $(1,2)$ asymmetric above. We show the positive frequency spectra for this example in Fig.~\ref{fig:spectra}, from which the symmetry and asymmetry of the different bipartitions can be seen.

\begin{figure}[tbhp]
\includegraphics[width=0.75\columnwidth]{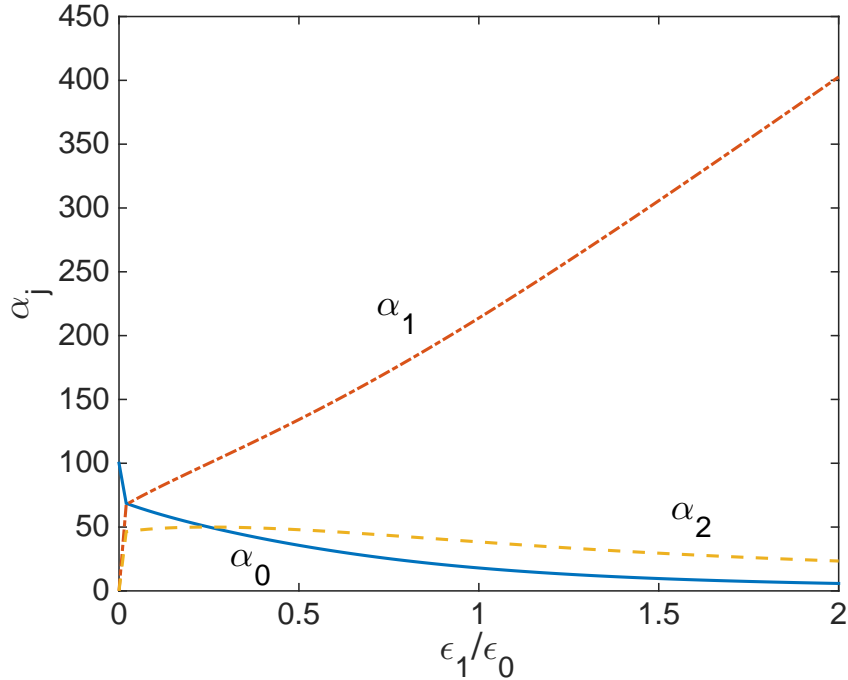}
\caption{(colour online) The steady-state mode amplitudes as a function of the ratio of the injected signal to the pump amplitude. $\gamma_{2}=\gamma_{0}=1=2\gamma_{1}$ for this result, $\kappa=10^{-2}$, and $\epsilon_{0}=100$.}
\label{fig:bright}
\end{figure}

In the normal nondegenerate OPO below threshhold, the downconverted fields have no coherent component and no fixed phase. This is no longer the case with an injected field at $\omega_{1}$, which sets a phase reference and thus gives a coherent component to both low frequency fields. In our case, we have treated both input fields as real and positive so that the intracavity fields in the resonant case are also real and positive. The fields have a bright coherent component, as can be seen in Fig.~\ref{fig:bright}, for $\gamma_{2}=\gamma_{0}=1=2\gamma_{1}$. Although not easily seen in the figure, $\alpha_{1}=\alpha_{2}=0$ for $\epsilon_{1}=0$ and in the non-injected case would maintain this value up to the oscillation threshhold.

In the three parameter regimes presented here, there is always at least one bipartition available which exhibits symmetric steering across the whole range $\epsilon_{1}/\epsilon_{c}$ that has been investigated. In Fig.~\ref{fig:EPRgequal} we find a region where only one bipartition has symmetric steering, and one where both do, although the degree of violation of the inequality by EPR$_{12}$ is small for larger $\epsilon_{1}$.  
In Fig.~\ref{fig:EPRg1g2two}, we see there is always one symmetric pair and one asymmetric pair, with these swapping roles at the same value of injected signal. The ability to choose the mode of operation adds flexibility to this scheme and may well have practical applications, beyond being of fundamental interest.



In conclusion, we have proposed a versatile and simple to operate means of producing tuneable symmetric and asymmetric steering between either modes of similar frequencies, or modes which are up to one octave apart in frequency. Optical parametric oscillators are mature technology, as is homodyne detection. The addition of a controllable input signal to an operating OPO is simplified by the fact that pumping lasers are often the result of frequency doubling from another laser output, meaning that a field at the appropriate frequency is already available. The control of the steering direction in this scheme is inherent to the apparatus itself, and does not depend on noise being added after the nonlinear interaction, as in previous proposals.

\end{document}